# Development of a Josephson junction based single photon microwave detector for axion detection experiments

D Alesini[1], D Babusci[1], C Barone[2,3], B Buonomo[1], M M Beretta[1], L Bianchini[4,5], G Castellano[6], F Chiarello[1,6], D Di Gioacchino[1], P Falferi[7,8], G Felici[1], G Filatrella[3,9], L G Foggetta[1], A Gallo[1], C Gatti[1], F Giazotto[4,10], G Lamanna[4,5], F Ligabue[4], N Ligato[4], C Ligi[1], G Maccarrone[1], B Margesin[8,11], F Mattioli[1,6], E Monticone[12,13], L Oberto[8,12], S Pagano[2,3], F Paolucci[12,13], M Rajteri[12,13], A Rettaroli[1,14], L Rolandi[4], P Spagnolo[4], A Toncelli[4,5], G Torrioli[1,6]

[1] INFN, Laboratori Nazionali di Frascati, Frascati, Roma, Italy
[2] Dipartimento di Fisica, Università di Salerno, Salerno, Italy
[3] INFN, Gruppo Collegato di Salerno, Salerno, Italy
[4] INFN, Sezione di Pisa, Pisa, Italy
[5] Dipartimento di Fisica, Università di Pisa, Pisa, Italy
[6] Istituto di Fotonica e Nanotecnologie CNR, Roma, Italy
[7] Istituto di Fotonica e Nanotecnologie CNR - Fondazione Bruno Kessler, Povo, Trento, Italy
[8] INFN, TIFPA, Povo, Trento, Italy
[9] Dipartimento di Scienze e Tecnologie, Università del Sannio, Salerno, Italy
[10] NEST, Pisa, Italy
[11] Fondazione Bruno Kessler, Povo, Trento, Italy
[12] Istituto Nazionale di Ricerca Metrologica (INRIM), Torino, Italy
[13] INFN, Sezione di Genova, Genova, Italy
[14] Dipartimento di Fisica, Università di Roma Tre, Roma, Italy

spagano@unisa.it

**Abstract**. Josephson junctions, in appropriate configurations, can be excellent candidates for detection of single photons in the microwave frequency band. Such possibility has been recently addressed in the framework of galactic axion detection. Here are reported recent developments in the modelling and simulation of dynamic behaviour of a Josephson junction single microwave photon detector. For a Josephson junction to be enough sensitive, small critical currents and operating temperatures of the order of ten of mK are necessary. Thermal and quantum tunnelling out of the zero-voltage state can also mask the detection process. Axion detection would require dark count rates in the order of 0.001 Hz. It is, therefore, is of paramount importance to identify proper device fabrication parameters and junction operation point.

## 1. Motivation

The low-mass frontier of Dark Matter, the measurement of the neutrino mass, the search for new light bosons in laboratory experiments, all require detectors sensitive to excitations of meV or smaller. Faint and rare signals, such as those produced by vacuum photoemission or by an Axion in a magnetic field, could be efficiently detected only by a new class of sensors. An increasing number of experiments intend to exploit the conversion of Dark Matter axions by means of strong magnetic fields into microwave photons. A comprehensive review of the current approaches to the detection of axion-like particles can be found in [1]. These experiments would benefit from the development of ultrasensitive microwave photon detectors. Additionally, low dark-count is a key requirement, as the typical expected rate of the axion-to-photon conversion is around $10^{-2} - 10^{-3}$ Hz [2].The Italian institute for nuclear physics (INFN) has financed the three-years project (SIMP) in order to strengthen its skills and technologies in this field, with the aim of developing a single microwave photon detector [3]. This goal will be pursued by improving the sensitivity and the dark count rate of two types of photodetectors: Josephson Junctions and Transition Edge Sensors.

## 2. Transition Edge Sensors

Transition Edge Sensors are the most sensitive photon and particle detectors available from THz to x-rays. The challenge is to extend their frequency (energy) sensitivity to the GHz range. In figure 1 is shown the principle of operation of a TES, for a detailed review see [4].

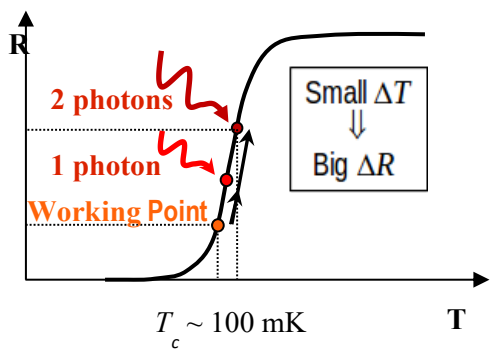

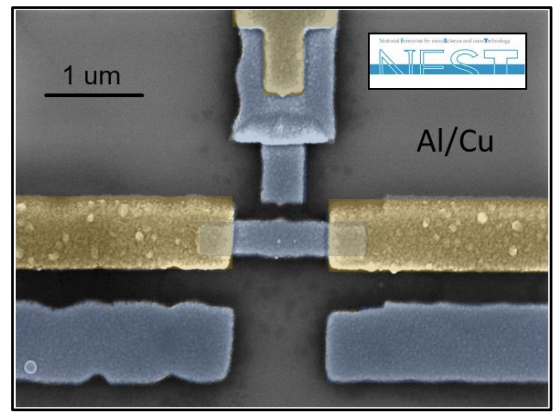

**Figure 1.** Detection principle of a TES. The red dots show the modification of the working point after the photon absorption.

**Figure 2.** Realization of a micro-TES to reduce heat capacity. Courtesy of NEST-Italy.

The TES energy resolution can be expressed as:

$$\Delta E \cong 2.35 \sqrt{2\,k_B\,T^2\,\frac{C}{\alpha}} \qquad (1)$$

where $k_B$ is the Boltzmann constant, $T$ the active region temperature, $C = \gamma\,V_a\,T$ the heat capacity, $\gamma$ the Sommerfeld coefficient, $V_a$ the active region volume, $\alpha = \frac{T}{R}\frac{dR}{dT}$ the figure of merit, and $R(T)$ the active region resistance. The directions of improvement to achieve a single microwave photon detection are:

- Improve $\alpha$
- Reduce $Tc$ by proximity effect or by current injection
- Reduce the heat capacity

A significant reduction of heat capacity can be obtained by scaling the TES size to sub-micron scale. In figure 2 it is shown an example of realization of a micro-TES. In addition, an increase of $\alpha$ and a decrease of $T_c$ can be achieved by proper choosing the composition of a S/N bilayer. In figure 3 are shown several $R$ vs $T$ curves obtained by changing the Ti/Au ratio and deposition parameters of a Ti/Au bilayer. As it is evident in the figure, there is ample space for improvement.

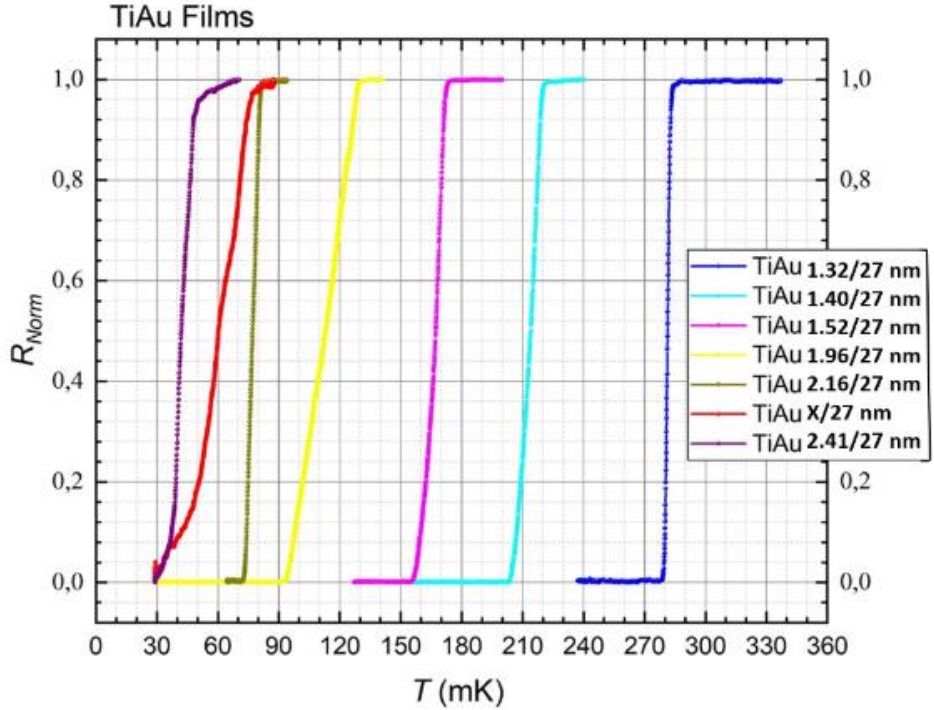

**Figure 3.** R vs T of various TiAu films, for different Ti/Au ratios (see legend), showing the possibility to engineer $T_c$ and $\alpha$. The Ti thickness corresponding to the red curve is not reported because it is not regular.

When all the optimization will be performed, a resolving power of: $h\nu/\Delta E \sim 6$–20 for 30–100 GHz photons, $V_a \approx 10^6\,\text{nm}^3$ and $T_c \approx 40$ mK can be expected.

Of course, sensitivity alone is not enough. Additional issues to be faced are an efficient coupling to the incoming photons and proper shielding from background radiation that could lead to significant dark counts. All these aspects are being actively investigated in the framework of the SIMP collaboration.

## 3. Josephson junctions

Josephson junctions, in appropriate configurations, can be excellent candidates for detection of single photons in the microwave frequency band. Such possibility has been recently addressed in the framework of galactic axion detection [5]. For a Josephson junction to be sensitive to a single microwave photon, whose energy is of the order of $10^{-24}$ J, are necessary operating temperatures of the order of few tens of mK. Moreover, thermal and/or quantum tunneling out of the zero-voltage state can mask the detection process. Axion detection would require dark count rates in the order of $10^{-3}$ Hz. It is, therefore, of paramount importance to identify proper device fabrication parameters and junction operation point.

### 3.1. Principle of operation

When a Josephson junction is biased by a dc current $I_{dc}$ below its critical current $I_0$, its phase difference $\phi$ stays in the minimum of a nonlinear potential $U(\phi)$, leading to a zero average voltage state. Such potential $U(\phi)$, defined as:

$$U(\phi) = I_0 \frac{\Phi_0}{2\pi}\left(1 - \cos(\phi) - \frac{I_{dc}}{I_0}\,\phi\right) \qquad (2)$$

where $\Phi_0$ is the flux quantum, is shown in figures 4 and 5.

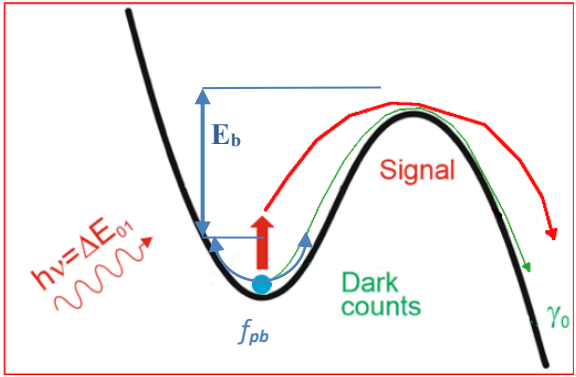

**Figure 4.** Thermal switching picture.

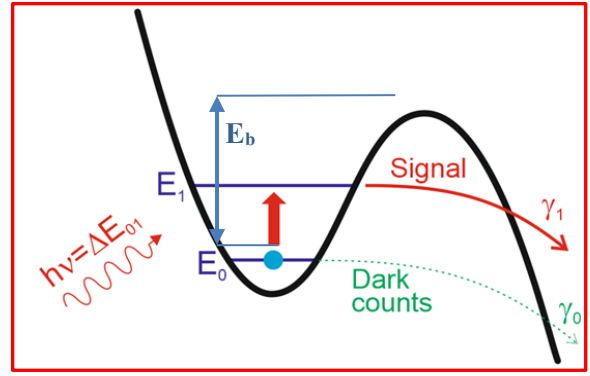

**Figure 5.** Quantum switching picture.

Separated by a current dependent energy barrier $E_b$, there is a "running" state where $\phi$ advances constantly in time, leading to a finite average voltage. Here, $E_b$ is expressed as:

$$E_b = I_0 \frac{\Phi_0}{\pi} \left( \sqrt{1 - \left(\frac{I_{dc}}{I_0}\right)^2} - \frac{I_{dc}}{I_0} \, arccos \left(\frac{I_{dc}}{I_0}\right) \right)$$ (3)

The passage from the zero to the finite voltage state can be triggered by an external signal, leading to a detection event, but also by thermal fluctuations or quantum tunneling.

Within the framework of the standard RSJ model of a Josephson junction, the absorption of one photon of microwave frequency $f_{PH}$ corresponds to a current pulse injected in the junction. The exact temporal dependence of such current is not well defined, beside its frequency. However, in correspondence of the photon absorption, there must be an energy transfer to the junction corresponding to one energy quantum $E_{PH} = h \, f_{PH}$, where $h$ is the Planck constant. Under proper conditions, the effect of the current pulse will be to have the phase $\phi$ overcome the energy barrier $E_b$ and switch to the "running" state, where an easily detectable voltage develops across the junction. In figure 4 such situation is shown. In absence of external signal, the junction phase should stay in the potential minimum $\phi_{min} = sin^{-1}(\frac{I_{dc}}{I_0})$. However, the finite operating temperature introduces random fluctuations, usually modelled as a random gaussian white current noise. This implies that there is a finite probability of the occurrence of a thermally activated switching (thermal dark counts), as also shown in figure 4. The thermal activation rate can be estimated as:

$$\Gamma_t \approx a_t \, f_{pb} \, e^{-\frac{E_b}{k_B T}}$$ (4)

where $f_{pb}$ is the, bias dependent, frequency of small "plasma" oscillations of the phase around the potential minimum:

$$f_{pb}(I_{dc}) = \frac{1}{2\pi} \sqrt{\frac{2e}{\hbar} \frac{J_c}{C_s}} \sqrt[4]{1 - \left(\frac{I_{dc}}{I_0}\right)^2}$$ (5)

here $J_c$ is the critical current density, $C_s$ the specific capacitance and $\alpha_t$ is an amplitude parameter defined, for low damping, as:

$$a_t = \frac{4\,\alpha}{[1+\sqrt{1+\frac{\alpha\,Q\,k_B T}{1.8\,E_b}}]^2} \qquad (6)$$

where $\alpha \approx 1$, and $Q = 2\,\pi\,f_{pb}(0)\,R\,C$ is the junction quality factor, with $R$ and $C$ the junction resistance and capacitance, respectively [6]. In the case we are interested, low temperatures and high-quality Al based junctions, the damping is very small and $Q$ is large, of the order of 1000.

At very low temperatures, it is expected that quantum aspects in the dynamics of the phase $\phi$ become important. In such quantum limit a switching through macroscopic quantum tunnelling can occur at a rate that, at zero temperature and in the absence of dissipation, can be calculated in the Wentzel-Kramers-Brillouin (WKB) approximation and for a cubic potential [6]:

$$\Gamma_q \approx a_q\,f_{pb}\,e^{-7.2\frac{E_b}{h\,f_{pb}}} \qquad (7)$$

where

$$a_q = \sqrt{120\,\pi\,7.2\,\frac{E_b}{h\,f_{pb}}} \qquad (8)$$

realizing the so-called quantum dark counts.

Within the quantum picture, the potential well hosts quantized energy levels, as shown in figure 5. The absorption of a microwave photon may induce the transition from a lower ($E_0$) to a higher ($E_1$) energy level, where the quantum tunnelling probability is much higher, thus triggering a switching phenomenon. In such case it is crucial that the photon energy $E_{ph}$ matches the energy level difference $\Delta E_{01} = E_1 - E_0$.

### 3.2. Junction design and operation points

The choice of the junction technology (e.g. Al/AlOx/Al) determines the value of $C_s$. In particular, for Al-based junctions realized by shadow evaporation technique, a value of $C_s \sim 100 \pm 25$ fF/$\mu$m$^2$ is reported [7]. Other parameters, such as the junction area $A_j$ and the critical current density Jc are defined at the fabrication stage, although the effective $J_c$ can be changed during the experiment by applying a suitable dc magnetic field. Conversely, the bias current $I_{dc}$ is not a free parameter. The axion detection experiment results in a single photon generated in a high-Q microwave cavity and therefore of a well-defined frequency. General considerations on the energy of the incoming axion and experimental constraints indicate a useful frequency range between 10 and 20 GHz [3]. To optimize the absorption of the photon by the junction, the effective Josephson plasma frequency $f_{pb}$ must then correspond to $f_{PH}$. By means of equation 5, this condition fixes the value of the bias current to:

$$I_{dc} = J_c\,A_j\,\sqrt{1 - \left(\frac{\hbar}{2e}\,C_s\,f_{PH}^{\,2}\,4\,\frac{\pi^2}{J_c}\right)^2} \qquad (9)$$

Although, considering the nonlinearity of the potential well, some small bias detuning may help to improve the detection sensitivity. $A_j$ and $J_c$ can then be properly chosen to optimize the junction sensitivity. However, as in the axion detection experiments the expected photon rate is very low (about 0.001 Hz), it is necessary to have comparatively small thermal and quantum escape rates from the potential minimum.

As an example, taking $A_j = 4\ \mu m^2$ and $J_j = 50\ A/cm^2$, the following dark count rates can be computed: $\Gamma_q \approx 0.002$ Hz and $\Gamma_t \approx 0.0002$ Hz, assuming $T = 0.11$ K. These rates are compatible with the expected axion detection rate ($\Gamma_{axion} \approx 0.001$ Hz) [2] and, by proper tuning the junction (area and critical current density) and experiment (temperature and local DC magnetic field) parameters, a favourable detection condition could be achieved.

By modelling the arrival of a single photon, with frequency $f_{PH}$ = 14GHz, as an oscillating current with a gaussian envelope, with a $FWHM$ = 450 ps and an amplitude of $I_{phmax}$ = 330 nA, a switching to the voltage state is observed, as shown in figures 6 and 7.

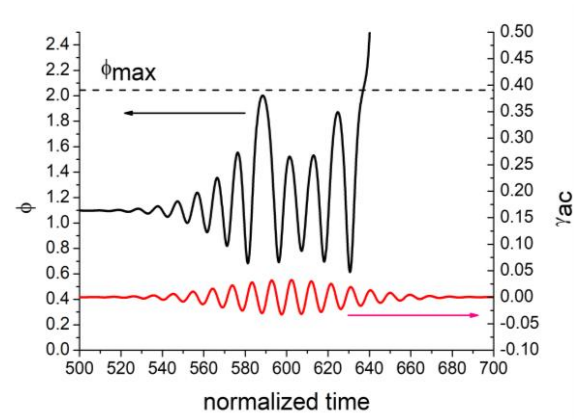

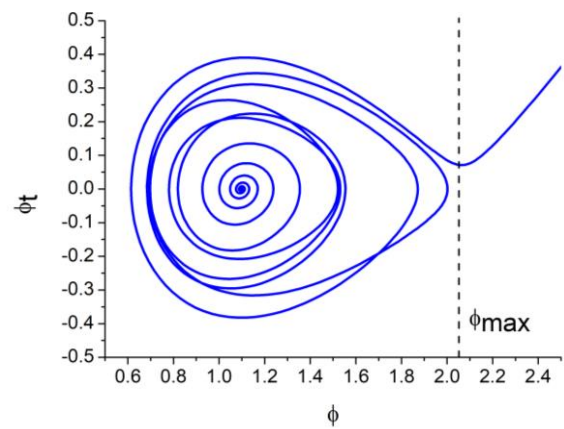

**Figure 6.** Simulation of the pulse induced switching. The single photon is simulated by a gaussian shaped oscillating current $I_{ph}$. The switching occurs when junction phase $\phi$ crosses $\phi_{max}$. The horizontal axis is the normalized time, while the vertical axis is the junction phase (black curve) and the normalized current (red curve).

**Figure 7.** Phase space diagram of the switching dynamics. On the horizontal axis is the junction phase $\phi$, while on the vertical axis is its time derivative $\dot{\phi}$. $\phi_{max}$ is the maximum value of the phase after which the junction switches to the voltage state.

Further work is in progress to extensively explore the junction parameters space and to identify the best operating points, where the sensitivity is maximized while keeping very low dark counts.

Moreover, the design of the RF circuit to effectively couple a single microwave photon to a Josephson junction having very small geometrical size, is being actively developed.

### 3.3. Device fabrication

Regarding the Al/AlO$_x$/Al Josephson junction fabrication, a shadow evaporation technique is being used, in order to realize the whole device without breaking the vacuum of the deposition chamber. This is necessary to assure a high junction quality (low quasiparticle losses). In figure 8 a sketch of the fabrication sequence is shown, with the relevant fabrication parameters indicated. In figure 9 an atomic force microscopy scan of one of the realized devices is shown, where the z scale allows to evaluate the low level of surface roughness achieved.

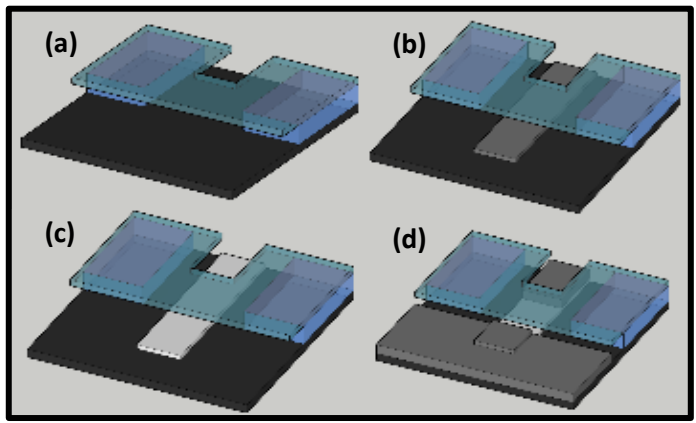

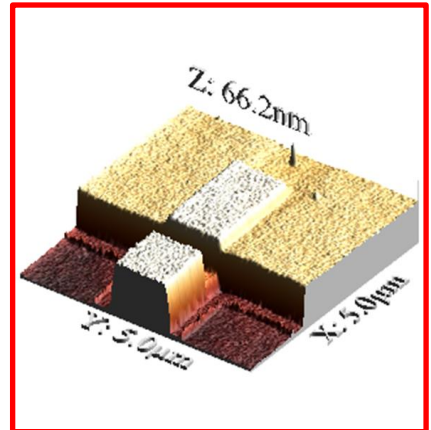

**Figure 8.** Fabrication steps of an Al/AlOₓ/Al Josephson junction using the shadow evaporation technique. (a) Photoresist bridge (blue and green structures); (b) First angled Al layer deposition (30 nm at $T$ = 155 °C); (c) Al oxidation ($t$ = 5 min, $P$ = 1 mbar); (d) Second angled Al layer deposition (30 nm at $T$ = 90 °C).

**Figure 9.** Atomic Force Microscopy scan of an Al/AlOx/Al Josephson Junction.

## 4. Conclusions

The current state of development of a microwave photon detector suitable to be used in Dark Matter experiments, such as the detection of axions, is reported. Currently specially designed TES and Josephson junctions are being developed to fulfill the sensitivity and dark count rate requirements. Specific considerations regarding the Josephson junction detector design have been reported.

## Acknowledgments

The authors wish to acknowledge financial support from Italian National Institute for Nuclear Physics INFN through the Project SIMP and from University of Salerno through projects FARB16CAVAL and FARB17PAGAN.

## References

[1]  Irastorza I G, Redondo J 2018 New experimental approaches in the search for axion-like particles *Prog. Part. Nucl. Phys.* **102** 89–159

[2]  Barbieri R, Braggio C, Carugno G, Gallo C S, Lombardi A, Ortolan A, Pengo R, Ruoso G and Speake C C 2017 Searching for galactic axions through magnetized media: The QUAX proposal *Phys. Dark Universe* **15** 135–141

[3]  Alesini D, *et al.* 2020 Status of the SIMP Project: Toward the Single Microwave Photon Detection *J. Low Temp. Phys.* **199** 348–354

[4]  Cabrera B 2008 Introduction to TES Physics *J. Low Temp. Phys.* **151** 82–93

[5]  Kuzmin L, Sobolev A S, Gatti C, Di Gioacchino D, Crescini N, Gordeeva A and Il'ichev E 2018 Single Photon Counter Based on a Josephson Junction at 14 GHz for Searching Galactic Axions *IEEE Trans. Appl. Supercond.* **28** 2400505

[6]  Martinis J M, Devoret M H and Clarke J 1987 Experimental tests for the quantum behavior of a macroscopic degree of freedom:The phase difference across a Josephson junction *Phys. Rev. B* **10** 4682-4698

[7]  Deppe F, Saito S, Tanaka H and Takayanagi H 2004 Determination of the capacitance of nm scale Josephson junctions *J. Appl. Phys.* **95** 2607-2613